\documentclass[11pt]{article}
\usepackage{epsf}
\usepackage{latexsym}
\begin{document} 
\newcommand{\eqnn}[1]{\begin{eqnarray*}#1\end{eqnarray*}}
\newcommand{\eqnl}[2]{\par\parbox{14cm}
{\begin{eqnarray*}#1\end{eqnarray*}}\hfill
\parbox{1cm}{\begin{eqnarray}\label{#2}\end{eqnarray}}}
\renewcommand{\baselinestretch}{1.3}
\newcommand{\eqnlb}[2]{\begin{equation}\fbox{$\displaystyle
#1 $}\label{#2}\end{equation}}
\newcommand{\eqngrlb}[3]{\par\parbox{12.5cm}
{\begin{eqnarray}\fbox{$\displaystyle   #1\\#2$}\end{eqnarray}}\hfill
\parbox{1cm}{\begin{eqnarray}\label{#3}\end{\eqnarray}}}
\newcommand{\eqngrl}[3]{\par\parbox{14cm}
{\begin{eqnarray*}#1\\#2\end{eqnarray*}}\hfill
\parbox{1cm}{\begin{eqnarray}\label{#3}\end{eqnarray}}}
\newcommand{\eqngr}[2]{\begin{eqnarray*}#1\\#2\end{eqnarray*}}
\newcommand{\eqngrr}[3]{\begin{eqnarray*}#1\\#2\\#3\end{eqnarray*}}
\newcommand{\eqngrrr}[4]{\begin{eqnarray*}#1\\#2\\#3\\#4\end{eqnarray*}}
\newcommand{\eqngrrrr}[5]{\begin{eqnarray*}#1\\#2\\#3\\#4\\#5
\end{eqnarray*}}
\newcommand{\eqngrrl}[4]{\par\parbox{12cm}
{\begin{eqnarray*}#1\\#2\\#3\end{eqnarray*}}\hfill
\parbox{1cm}{\begin{eqnarray}\label{#4}\end{eqnarray}}}
\newcommand{\casegr}[5]{\[#1 =\Bigg\{\begin{array}{cl}
#2 & #3\\ #4 & #5 \end{array}\]}
\newcommand{\eqngrrrl}[5]{\par\parbox{12cm}
{\begin{eqnarray*}#1\\#2\\#3\\#4\end{eqnarray*}}\hfill
\parbox{1cm}{\begin{eqnarray}\label{#5}\end{eqnarray}}}
\newcommand{\refs}[1]{(\ref{#1})}
\def\fd{{^{*\!}F}}
\def\intl{\int\limits}
\def\vn{\vec{n}}
\def\vxi{\vec{\xi}}
\def\vk{\vec{k}}
\def\vm{\vec{m}}
\def\nup{\vert n_\uparrow\rangle}
\def\mup{\vert m_\uparrow\rangle}
\def\ndown{\vert n_\downarrow\rangle}
\def\psiup{\psi_\uparrow}
\def\psidown{\psi_\downarrow}
\def\hha{{\hbar\ov 2}}
\def\un{\underline}
\def\tsi{\tilde\sigma}
\def\hb{\hbar}
\def\dett{{\det}_\theta}
\def\di{D\!\!\!\!\slash\,}
\def\as{a\!\!\!\slash\,}
\def\ps{p\!\!\!\slash\,}
\def\fdi{\partial\!\!\!\slash\,}
\def\ddx{d^dx}
\def\va{\vec{a}}
\def\vb{\vec{b}}
\def\vx{\vec{x}}
\def\vy{\vec{y}}
\def\vL{\vec{L}}
\def\vs{\vec{s}}
\def\vOmega{\vec{\Omega}}
\def\vtheta{\vec{\theta}}
\def\val{\vec{\alpha}}
\def\vp{\vec{p}}
\def\vv{\vec{v}}
\def\vV{\vec{V}}
\def\vsigma{\vec{\sigma}}
\def\vA{\vec{A}}
\def\vE{\vec{E}}
\def\vpi{\vec{\pi}}
\def\vB{\vec{B}}
\def\vj{\vec{j}}
\def\vM{\vec{M}}
\def\vS{\vec{S}}
\def\vJ{\vec{J}\,}
\def\ve{\vec{e}}
\def\hi{{\hbar\ov i}}
\def\ih{{i\ov \hbar}}
\def\nablad{\nabla\cdot}
\def\tr{\,{\rm tr}\,}
\def\bet{\bar\eta}
\def\bal{\bar\alpha}
\def\bps{\bar\psi}
\def\gan{\gamma_n}
\def\cJ{{\cal J}}
\def\cG{{\cal G}}
\def\cF{{\cal F}}
\def\cR{{\cal R}}
\def\cL{{\cal L}}
\def\cc{{\cal C}}
\def\cd{{\cal D}}
\def\cs{{\cal S}}
\def\ch{{\cal H}}
\def\ra{\rangle}
\def\la{\langle}
\def\gf{\gamma_5}
\def\gff{\bar\gamma}
\def\pan{\par\noindent}
\def\cov{\bigtriangledown}
\def\pd{\psi^{\dagger}}
\def\mtxt#1{\quad\hbox{{#1}}\quad}
\def\pa{\partial}
\def\gam{\gamma}
\def\eps{\epsilon}
\def\es{\!=\!}
\def\lapf{\triangle_f}
\def\lap{\triangle}
\def\olap{{1\ov\lap}}
\def\ov{\over}
\def\cd{{\cal D}}
\def\om{\omega}
\def\pamu{{\partial_\mu}}
\def\panu{{\partial_\nu}}
\def\al{\alpha}
\def\be{\beta}
\def\si{\sigma}
\def\sp{\hbox{Sp}\,}
\def\pr{\prime}
\def\lam{\lambda}
\def\tr{\hbox{tr}\,}
\def\cm{{\cal M}}
\def\cf{{\cal F}}
\def\cmb{{\pa{\cal M}}}
\def\ha{{1\over 2}}
\def\Schr{Schr\"odinger}
\def\Schrs{Schr\"odingers\,}
\def\cl{{\cal L}}
\title{Quantum Fields near Black Holes\footnote{to appear in
\textit{Black Holes: Theory and Observation}, edited by
F.W. Hehl, C. Kiefer and R. Metzler (Springer, Berlin, 1998).}}
\author{Andreas Wipf\\
Theoretisch-Physikalisches Institut,
Friedrich-Schiller-Universit\"at\\
Max Wien Platz 1, 07743 Jena}
\date{}
\maketitle
\def\cM{{\cal M}}
\def\cN{{\cal N}}
\def\cD{{\cal D}}
\def\cl{{\cal L}}
\def\cP{{\cal P}}
\def\cS{{\cal S}}
\def\cSC{{\cal S}^C}
\def\cH{{\cal H}}
\begin{abstract}This review gives an introduction into 
problems, concepts and techniques when quantizing
matter fields near black holes. The first part
focusses on quantum fields in general curved
space-times. The second part is devoted to
a detailed treatment of the Unruh effect in
uniformly accelerated frames and the Hawking
radiation of black holes. Particular emphasis is
put on the induced energy momentum tensor near black holes.
\end{abstract}

\section{Introduction}
In the theory of quantum fields on curved space-times
one considers gravity as a classical background
and investigates quantum fields propagating
on this background.
The structure of spacetime is described by a
manifold $\cM$ with metric $g_{\mu\nu}$.
Because of the
large difference between the Planck scale ($10^{-33}$cm)
and scales relevant for the present standard model
($\geq 10^{-17}$cm) the range of validity
of this approximation should include a wide variety 
of interesting phenomena, such as
particle creation near a black hole with Schwarzschild
radius much greater than the Planck length.

The difficulties in the transition from flat
to curved spacetime lie in the absence of the notion
of global inertial observers or of Poincar\'e 
transformations which underlie the concept of
particles in Minkowski spacetime. 
In flat spacetime, Poincar\'e symmetry is 
used to pick out a preferred irreducible representation
of the canonical commutation relations.
This is achieved by selecting an
invariant vacuum state and hence a particle notion. 
In a general curved spacetime
there does not appear to be any preferred concept of 
particles. If one 
accepts that quantum field theory
on general curved spacetime is a quantum theory of \textit{fields},
not particles, then the existence of global 
inertial observers is irrelevant for the formulation of 
the theory. For linear fields a satisfactory
theory can be constructed. 
Recently Brunelli and Fredenhagen \cite{Fredenhagen}
extended the Epstein-Glaser scheme to curved 
space-times (generalising
an earlier attempt by Bunch \cite{Bunch})
and proved perturbative
renormalizability of $\lam\phi^4$.

The framework and structure of Quantum field theory 
in curved space-times emerged from Parker's analysis
of particle creation in the very early universe
\cite{Parker}. The theory received enormous impetus
from Hawking's discovery that black holes radiate 
as black bodies due to particle creation \cite{Hawking}.
A comprehensive summary of the work
can be found in the books
\cite{Birrell}.
\section{Quantum Fields in Curved Spacetime}
In a general spacetime no analogue of a 'positive frequency
subspace' is available and as a consequence the states
of the quantum field will not possess a physically
meaningful particle interpretation. In addition,
there are spacetimes, e.g. those with time-like singularities, 
in which solutions of the wave equation cannot
be characterised by their initial values.
The conditions of \textit{global hyperbolocity}
of $(\cM,g_{\mu\nu})$ excludes such 'pathological'
spacetimes and ensures that
the field equations have a well posed
initial value formulation.
Let $\Sigma\subset\cM$ be a hypersurface
whose points cannot be joined
by time-like curves. We define the \textit{domain
of dependence of} $\Sigma$ by
\eqnn{
\hbox{D}(\Sigma)=\{p\in\cM\vert\hbox{every inextendible
causal curve through $p$ intersects } \Sigma\}.}
If D$(\Sigma)=\cM$, $\Sigma$ is called a 
\textit{Cauchy surface} for the spacetime and
$\cM$ is called \textit{globally hyperbolic}.
Globally hyperbolic spacetimes can be
\textit{foliated} by a one-parameter family of smooth
Cauchy surfaces $\Sigma_t$, i.e. a smooth 'time
coordinate' $t$ can be chosen on $\cM$ such that
each surface of constant $t$ is a Cauchy surface \cite{Geroch}.
There is a \textit{well posed initial
value problem} for linear wave equations \cite{HawEl}. 
For example, given smooth initial data
$\phi_0,\dot\phi_0$, then there
exists a unique solution $\phi$ of the \textit{Klein-Gordon
equation}
\eqnl{
\Box_g\phi+m^2\phi=0,\qquad \Box_g={1\ov \sqrt{-g}}\pa_\mu(\sqrt{-g}
g^{\mu\nu}\pa_\nu)}{KK}
which is smooth
on all of $\cM$, such that on $\Sigma$ we have
$\phi=\phi_0\mtxt{and}n^\mu\nabla_\mu\phi=\dot\phi_0,$
where $n^\mu$ is the unit future-directed normal to
$\Sigma$. In addition,  $\phi$
varies continuously with the initial data.
\pan
For the phase-space formulation
we slice $\cM$ by space-like
Cauchy surfaces $\Sigma_t$ and introduce unit normal
vector fields $n^\mu$ to $\Sigma_t$. The spacetime metric
$g_{\mu\nu}$ induces a spatial metric $h_{\mu\nu}$ on 
each $\Sigma_t$ by the formula
\eqnn{
g_{\mu\nu}=n_\mu n_\nu-h_{\mu\nu}.}
Let $t^\mu$ be a 'time evolution' vector field on $\cM$ 
satisfying $t^\mu \nabla_\mu t=1$. We decompose it into
its parts normal and tangential to $\Sigma_t$, 
\eqnn{
t^\mu=Nn^\mu+N^\mu,}
where we have defined the \textit{lapse function}  $N$ and 
the \textit{shift vector} $N^\mu$ tangential to the $\Sigma_t$.
Now we introduce adapted coordinates $x^\mu=(t,x^i), i=1,2,3$ with
$t^\mu\nabla_\mu x^i=0$, so that $t^\mu\nabla_\mu=\pa_t$
and $N^\mu\pa_\mu=N^i\pa_i$. The metric coefficients in
this coordinate system are
\eqnn{
g_{00}=g(\pa_t,\pa_t)=N^2-N^iN_i\mtxt{and}
g_{0i}=g(\pa_t,\pa_i)=-N_i,}
where $N_i=h_{ij}N^j$,
so that
\eqngr{
ds^2&=&(Ndt)^2-h_{ij}(N^idt+dx^i)(N^jdt+dx^j)}
{(\pa\phi)^2&=&{1\ov N^2}(\pa_0\phi-N^i\pa_i\phi)^2-
h^{ij}\pa_i\phi\pa_j\phi.}
The determinant $g$ of the $4$-metric is 
related to the determinant $h$ of the $3$-metric as $g=-N^2h$.
Inserting these results into the Klein-Gordon 
action
\eqnn{
S=\int L dt=\ha\int \eta\Big(g^{\mu\nu}\pa_\mu\phi\pa_\nu\phi
-m^2\phi^2\Big),\qquad \eta=\sqrt{\vert g\vert}d^4x,}
one obtains for the momentum density, $\pi$, conjugate
to the configuration variable $\phi$ on $\Sigma_t$
\eqnn{
\pi={\pa L\ov \pa\dot\phi}={\sqrt{h}\ov N}\big(
\dot\phi-N^i\pa_i\phi\big)=\sqrt{h}\big(n^\mu\pa_\mu\phi\big).}
A point in classical phase space
consists of the specification of functions
$(\phi,\pi)$ on a Cauchy surface. 
By the result of Hawking and Ellis, smooth $(\phi,\pi)$ 
give rise to a unique solution to \refs{KK}. 
The space of solutions is independent on the choice 
of the Cauchy surface.\par
For two (complex) solutions of the Klein-Gordon equation 
the inner product
\eqnn{(u_1,u_2)\equiv
i\intl_{\Sigma}\Big(\bar u_1 n^\mu\nabla_\mu u_2-
(n^\mu\nabla_\mu \bar u_1)u_2\Big)\sqrt{h}\,d^3x=
i\int\big(\bar u_1\pi_2-\bar\pi_1 u_2\big)d^3x}
defines a natural symplectic structure. 
Natural means, that $(u_1,u_2)$ is independent of the
choice of $\Sigma$.
This inner product is not positive definite.
Let us introduce a complete set of conjugate pairs
of solutions $(u_k,\bar u_k)$ of the
Klein-Gordon equation\footnote{the $k$ are
any labels, not necessarily the momentum} 
satisfying the following ortho-normality conditions
\eqnn{
(u_k,u_{k^\pr})=\delta(k,k^\pr)\Rightarrow
(\bar u_k,\bar u_{k^\pr})=-\delta(k,k^\pr)
\mtxt{and}
(u_k,\bar u_{k^\pr})=0.}
There will be an infinity of such sets. Now we expand
the field operator in terms of these modes:
\eqnn{
\phi=\int d\mu(k)\Big(a_k u_k+a_k^\dagger \bar u_k\Big)
\mtxt{and}
\pi=\int d\mu(k)\Big(a_k \pi_k+a_k^\dagger\bar\pi_k\Big),}
so that
\eqnn{
(u_k,\phi)=a_k\mtxt{and}(\bar u_k,\phi)=-a^\dagger_k.}
By using the completeness of the $u_k$ and the
canonical commutation relations
one can show that the operator-valued
coefficients $(a_k,a^\dagger_k)$ satisfy the
usual commutation relations
\eqnl{
[a_k,a_{k^\pr}]=[a^\dagger_k,a^\dagger_{k^\pr}]=0\mtxt{and}
[a_k,a^\dagger_{k^\pr}]=\delta(k,k^\pr).}{comrel}
We choose the Hilbert space $\cH$ to be the Fock space
built from a 'vacuum' state $\Omega_u$ satisfying
\eqnl{
a_k\Omega_u=0\mtxt{for all}k,\qquad (\Omega_u,\Omega_u)_{\cH}
=1.}{comrel1}
The 'vectors' $\Omega_u,a^\dagger_k\Omega_u,\dots$
comprise a basis of $\cH$. The scalar product
given by (\ref{comrel},\ref{comrel1}) is positive-definite.\pan
If $(v_p,\bar v_p)$ is a
second set of basis functions, we may
as well expand the field operator in terms of this
set
\eqnn{
\phi=\int d\mu(p)\Big(b_p v_p+
b_p^\dagger\bar v_p\Big).}
The second set will be linearly related to the first one by 
\eqnn{
v_p=\int d\mu(k)\Big((u_k,v_p)u_k
-(\bar u_k,v_p)\bar u_k\Big)\equiv
\int d\mu(k)\Big(\al(p,k)u_k+\beta(p,k)\bar u_k
\Big).}
The inverse transformation reads
\eqnn{
u_k=\int d\mu(p)\Big(v_p\bar\al(p,k)-\bar v_p\beta(p,k)\Big).}
As a consequence, the Bogolubov-coefficients are related by
\eqnl{
\al\al^\dagger-\beta\beta^\dagger=1\mtxt{and}
\al\beta^t-\beta\al^t=0.}{bogrel}
If the $\beta(k,p)$ vanish, then the 'vacuum' is left
unchanged, but if they do not, we have
a nontrivial \textit{Bogolubov transformation}
\eqnl{
\pmatrix{a&a^\dagger}=\pmatrix{b&b^\dagger}\pmatrix{\al&\beta\cr
\bar\beta&\bar\al}\mtxt{and}
\pmatrix{b\cr b^\dagger}=\pmatrix{\bar\al&-\bar\beta\cr
-\beta&\al}\pmatrix{a\cr a^\dagger}}{bogtrans}
which mixes the annihilation and creations operators.
If one defines a Fock space and a 'vacuum' corresponding
to the first mode expansion, $a_k\Omega_u=0$,
then the expectation of the number operator $b^\dagger_p b_p$ defined
with respect to the second mode expansion is
\eqnn{
\big(\Omega_u,b_p^\dagger b_p\Omega_u\big)
=\int d\mu(k)\vert \beta(p,k)\vert^2.} 
That is, the old vacuum contains new particles. It may
even contain an infinite number of new particles, in
which case the two Fock spaces cannot be related
by a unitary transformation.

\textbf{Stationary and static spacetimes.}
A spacetime is \textit{stationary} if there exist
coordinates for which the metric
is time-independent. 
This property holds iff spacetime admits
a time-like Killing field $K=K^\mu\pa_\mu$
and hence a natural choice for the mode functions $u_k$:
We may scale $K$ such that the Killing time $t$
is the proper time measured by at least one comoving
clock. Now we may choose as basis functions $u_k$ the
eigenfunctions of the Lie derivative,
\eqnn{
iL_Ku_k=\om(k)u_k\mtxt{and}iL_K\bar u_k=-\om(k)\bar u_k,}
where the $\om(k)>0$ are constant.
The $\om(k)$ are the frequencies relative to
the particular comoving clock and the $u_k$ and $\bar u_k$ 
are the positive and negative frequency solutions, 
respectively. Now the construction of the vacuum and
Fock space is done as described above.\pan
In a \textit{static spacetime}, $K$
is everywhere orthogonal to a family
of hyper-surfaces and hence
satisfies the Frobenius condition
$\tilde K\wedge d\tilde K=0,\quad \tilde K=K_\mu dx^\mu.$
We may introduce adapted coordinates:
$t$ along the congruence ($K=\pa_t$) and $x^i$ in one hypersurface
such that the metric is time-independent
and the shift vector $N_i$ vanishes,
\eqnn{
(g_{\mu\nu})=\pmatrix{N^2(x^i)&0\cr 0&-h_{ij}(x^i)}.}
As modes we use
\eqnn{
u_k={1\ov \sqrt{2\om(k)}}e^{-i\om(k)t}\phi_k(x^i)}
which diagonalise $L_K$ and for which
the Klein-Gordon equation simplifies to
\eqnn{{\cal K}\phi_k\equiv
\Big(-{N\ov \sqrt{h}}\pa_i\big(N\sqrt{h}
h^{ij}\pa_j\big)+N^2m^2\Big)\phi_k=\om_k^2\phi_k.}
Since $n^\mu\pa_\mu =N^{-1}\pa_t$,
the inner product of two mode functions is
\eqnn{
(u_1,u_2)={\om_1+\om_2\ov 2\sqrt{\om_1\om_2}}\;e^{i(\om_1-\om_2)t}
\underbrace{\int \bar\phi_1\phi_2\;N^{-1}\sqrt{h}\,d^3x}_{(\phi_1,
\phi_2)_2}.}
The elliptic operator ${\cal K}$ is symmetric 
with respect to the $L_2$ scalar product $(.,.)_2$
and may be diagonalised. Its positive eigenvalues are the
$\om^2(k)$ and its
eigenfunctions form a complete 'orthonormal' set on $\Sigma$,
$(\phi_k,\phi_{k^\pr})_2=\delta(k,k^\pr)$. It follows then
that the $u_k$ form a complete set with the properties
discussed earlier.

Ashtekar and Magnon \cite{ashmag}
and Kay \cite{kay} gave a rigorous construction
of the Hilbert space and Hamiltonian in a stationary spacetime.
They started with a \textit{conserved positive scalar product} $(.,.)_E$
\eqnn{
(\phi_1,\phi_2)_E=\intl_\Sigma T_{\mu\nu}(\phi_1,\phi_2)
K^\nu n^\mu\sqrt{h}d^3x,}
where the bilinear-form on the space of complex solutions
is defined by the metric 'stress tensor': 
\eqnn{
T_{\mu\nu}(\phi,\psi)
=\ha\Big(\phi^\dagger,_\mu\psi,_\nu+\phi^\dagger,_\nu\psi,_\mu
-g_{\mu\nu}\big(\nabla\phi^\dagger \nabla\psi-
m^2\phi^\dagger\psi\big)\Big).}
This 'stress tensor' is symmetric and conserved 
and hence $\nabla_\mu(T^{\mu\nu}K_\nu)=0$. 
It follows that the norm is invariant
under the time-translation map
\eqnn{
\al_t^*(\phi)=\phi\circ\al_t\mtxt{or}
\big(\al_t^*(\phi)\big)(x)=\phi\big(\al_t(x)\big),}
generated by the Killing field $K$. When completing the 
space of complex solutions in the 'energy-norm'
one gets a complex (auxiliary) Hilbert space $\tilde \cH$. 
The time translation map extends to $\tilde\cH$ and defines
a one-parameter unitary group
\eqnn{
\al^*_t=e^{i\tilde ht},\qquad \tilde h\mtxt{self-adjoint.}}
Note, that from the definition of the Lie derivative,
\eqnn{
{d\ov dt}\big(\al^*_t\phi\big)\vert_{t=0}=-L_K\phi=
i\tilde h\phi.}
The conserved inner product
$(\phi_1,\phi_2)$
can be bounded
by the energy norm and hence extends to a quadratic
form on $\tilde\cH$.
Let $\tilde {\cH}^+\subset \tilde \cH$ be the positive spectral subspace
in the spectral decomposition of $\tilde h$ 
and let $P$ be the projection map
$P:\tilde\cH\to \tilde\cH^+$. For all real solutions
we may now define the \textit{scalar product} as
the inner product of the projected solutions, which
are complex. The one-particle Hilbert space $\cH$ is just the completion
of the space $\tilde \cH^+$ of 'positive frequency solutions'
in the Klein-Gordon inner product.\pan
\textbf{Hadamard states.}
For a black hole the global Killing field is not 
everywhere time-like. One
may exclude the non-time-like region from space time
which corresponds to the imposition of boundary
conditions. One may also try to retain
this region but attempt to define
a meaningful vacuum by invoking physical arguments.
In general spacetimes there is no Killing vector at all.
One probably has to give up the particle
picture in this generic situation.

In (globally hyperbolic) spacetimes without any
symmetry one can still construct a well-defined
Fock space over a quasifree vacuum state,
provided that the two-point functions satisfies the 
so-called Hadamard condition.
Hadamard states are states, for which the two-point
function has the following singularity structure
\eqnl{
\om\big(\phi(x)\phi(y)\big)\equiv
\om_2(x,y)={u\ov\sigma}+v\log\sigma +w,}{hadamard}
where $\sigma(x,y)$ is the square of the geodesic distance
of $x$ and $y$ and $u,v,w$ are smooth functions on
$\cM$. It has been shown 
that if $\om_2$ has the Hadamard singularity structure
in a neighbourhood of a Cauchy-surface, then it has
this form everywhere \cite{fullingsweeny}. To show that,
one uses that $\om_2$ satisfies the wave equation.
This result can then be used to show that on a
globally hyperbolic spacetime there is a wide class of
states whose two-point functions have the Hadamard singularity
structure.\pan
The two-point function $\om_2$ must be positive,
\eqnn{
\om \big(\phi(f)^\dagger \phi(f)\big)=
\int d\mu(x)d\mu(y)\;\bar f(x)\om_2\big(x,y\big)f(y)\geq 0,}
and must obey the Klein-Gordon equation. 
These requirements determine $u$ and $v$ uniquely and put
stringent conditions on the form of $w$.
In a globally hyperbolic spacetime there are unique retarded and advanced
Green functions
\eqnn{
\Delta_{ret}(x,y)\mtxt{,} \Delta_{adv}(x,y)\mtxt{with }
\hbox{ supp}(\Delta_{ret})=\{(x,y);x\in J_+(y)\},}
where $J_+(y)$ is the causal future of $y$.
The \textit{Feynman Green function} is related
to $\om_2$ and the advanced Green function as
\eqnn{
i\Delta_F(x,y)=\om_2(x,y)+\Delta_{adv}(x,y).}
Since $\Delta_{adv}$ is unique,
the ambiguities of $\Delta_F$ are the same as those of
$\om_2$. The \textit{propagator function}
\eqnn{
i\Delta(x,y)=[\phi(x),\phi(y)]=\Delta_{ret}(x,y)-\Delta_{adv}(x,y)}
determines the antisymmetric part of $\om_2$,
\eqnn{
\om_2(x,y)-\om_2(y,x)=i\Delta(x,y),}
so that this part is without ambiguities.
For a scalar field without self-interaction we expect that
\eqnn{
\om\big(\phi(x_1)\dots\phi(x_{2n-1})\big)=0,\qquad
\om\big(\phi(x_1)\dots\phi(x_{2n})\big)=\sum_{i_1<i_2\dots <i_n\atop
j_1<j_2\dots <j_n}\prod_{k=1}^n\om\big(
\phi(x_{i_k})\phi(x_{j_k})\big).}
A state $\om$ fulfilling these conditions is called
\textit{quasifree}.
Now one can show that any choice of $\om_2(x,y)$
fulfilling the properties listed above gives rise to a well-defined
Fock-space $\cF=\oplus \cF_n$
over a quasifree vacuum state. 
The scalar-product on the 'n-particle subspace'
$\cF_n$ in
\eqnl{
\cF_n=\{\psi\in \cD(\cM^n)_{symm}/\cN\}^{completion}
,\qquad n=0,1,2,\dots,}{hilbertn}
where $\cD(\cM^n)$ denotes the smooth symmetric functions on
$\cM\times \cdots\times \cM$ ($n$ factors) with compact support, 
is
\eqnn{
(\psi_1,\psi_2)=\int d\mu(x_1,..,x_n,y_1,..,y_n)
\prod_{i=1}^n\om_2(x_i,y_i)\bar\psi_1(x_1,..,x_n)\psi_2(y_1,..,
y_n),}
where $d\mu(x_1,x_2,..)=
d\mu(x_1)d\mu(x_2)\dots$. Since $\om_2$ satisfies the
wave equation,
the functions in the image of $\Box+m^2$ have zero norm. The 
set of zero-norm states $\cN$ has been
divided out in order to end up with a
positive definite Hilbert space.\pan
The smeared field operator is now defined in the usual way:
$\phi(f)=a(f)^\dagger+a(\bar f)$,
where
\eqngr{
\big(a(\bar f)\psi\big)_n(x_1,..,x_n)&=&
\sqrt{n+1}\int d\mu(x,y)\om_2(x,y)f(x)\psi_{n+1}(y,x_1,..,x_n)}
{\big(a(f)^\dagger\psi\big)_n(x_1,..,x_n)&=&{1\ov\sqrt{n}}
\sum\limits_{k=1}^n
f(x_k)\psi_{n-1}(x_1,..,x_{k-1},x_{k+1}..,x_n),\quad n>0}
and $(a(f)^\dagger\psi)_0=0$.
It is now easy to see that $\om_2$ is just the Wightman function
of $\phi$ in the vacuum state $\psi_0$:
$\om_2(x,y)=\big(\psi_0,\phi(x)\phi(y)\psi_0\big)$.
\section{The Unruh Effect}
We may ask the question how quantum fluctuations
appear to an accelerating observer? In particular,
if the observer was carrying with him a robust detector,
what would this detector register?
If the motion of the observer undergoing
constant (proper) acceleration is confined 
to the $x^3$ axis, then
the world line is a hyperbola in the $x^0,x^3$ plane
with asymptotics $x^3=\pm x^0$. These asymptotics
are \textit{event horizons} for the accelerated observer.
To find a natural comoving frame we
consider a family of accelerating observers,
one for each hyperbola with asymptotics $x^3=\pm x^0$.
The coordinate system is then the comoving
one in which along each hyperbola the space coordinate
is constant while the time coordinate $\tau$ is proportional
to the proper time as measured from an
initial instant $x^0=0$ in some inertial frame.
The world lines of the uniformly accelerated particles are
the orbits of one-parameter group of Lorentz boost isometries
in the $3$-direction:
\eqnn{
\pmatrix{x^0\cr x^3}=\rho\pmatrix{\sinh \kappa t\cr \cosh \kappa t}
=e^{\kappa\om t}\pmatrix{0\cr \rho},
\qquad(\om^\mu_{\,\nu})=\pmatrix{0&1\cr 1&0}.}
In the comoving coordinates $(t,x^1,x^2,\rho)$
\eqnn{
ds^2=\kappa^2\rho^2 dt^2-d\rho^2-(dx^1)^2-(dx^2)^2.} 
so that the proper time along a hyperbola $\rho=$const
is $\kappa\rho t$.
The orbits are tangential to the \textit{Killing field}
\eqnl{
K=\pa_t=\kappa(x^3\pa_0+x^0\pa_3)\mtxt{with}(K,K)=(\kappa\rho)^2=g_{00}.}{kill}
Some typical orbits are depicted in figure \refs{rindler1}.
\begin{figure}[ht]
\begin{minipage}[t]{15cm}
\centerline{\epsfysize=7 cm\epsffile{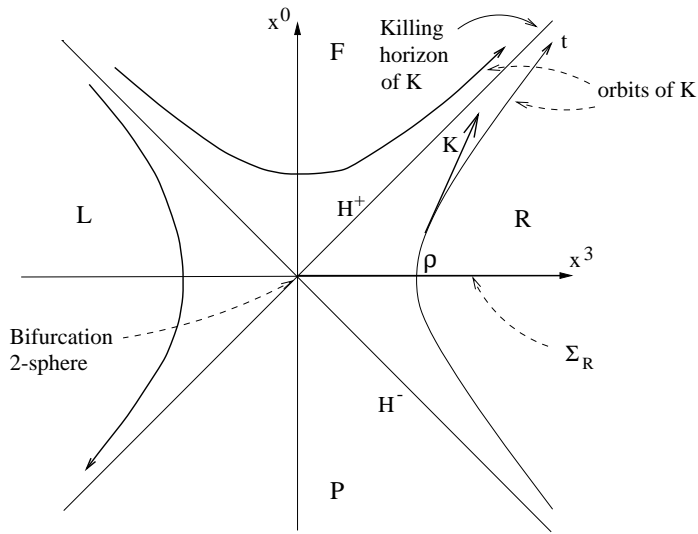}}
\caption{\label{rindler1}\textsl{A
Rindler-observer sees only a quarter of Minkowski
space}}
\end{minipage}
\end{figure}
Since the proper acceleration on the orbit with
$(K,K)=1$ or $\rho=1/\kappa$ is $\kappa$, it is
conventional to view the orbits of $K$ as corresponding
to a family of observers associated with an observer
who accelerates uniformly with acceleration $a=\kappa$.\pan
The coordinate system $t,\rho$ covers the Rindler
wedge $R$ on which $K$ is time-like future directed.
The boundary $H^+$ and $H^-$ of the wedge 
is given by $\rho=0$ and appears as a 
\textit{Killing horizon}, on which $K$ becomes null. 
Beyond this event horizon the Killing vector field 
becomes space-like in the regions
$F,P$ and time-like past directed in $L$.
The parameter $\kappa$ plays the role of the \textit{surface
gravity}. To see that, we set $r-2M=\rho^2/8M$
in the Schwarzschild solution
and linearise the metric near the horizon $r\sim 2M$.
One finds that
\eqnn{
ds^2\sim\underbrace{
(\kappa \rho)^2dt^2-d\rho^2}_{\hbox{\tiny 2-dim Rindler}\atop
 \hbox{\tiny spacetime}}
-\underbrace{{1\ov 4\kappa^2}d\Omega^2}_{\hbox{\tiny 2-sphere of} \atop
\hbox{\tiny radius }1/2\kappa}}
contains the line element of two-dimensional Rindler spacetime,
where $\kappa=1/4M$ is indeed the surface gravity of the Schwarzschild
black hole.\pan
\textbf{Killing horizons
and surface gravity.}
The notion of Killing horizons is relevant
for the Hawking radiation and the thermodynamics
of black holes and can already be illustrated in
Rindler spacetime.
Let $S(x)$ be a smooth function
and consider a family of hyper-surfaces $S(x)=\,$const.
The vector fields normal to the hyper-surfaces are
\eqnn{
l=g(x)(\pa^\mu S)\pa_\mu,}
with arbitrary non-zero function $g$. If $l$ is null, 
$l^2=0$, for a
particular hypersurface $\cN$ in the family,
$\cN$ is said to be a \textit{null hypersurface}.
For example, the normal vectors to the
surfaces $S=r-2M=\,$const in Schwarzschild spacetime have norm
\eqnn{
l^2=g^2 g^{\mu\nu}\pa_\mu S\pa_\nu S=g^2\Big(1-{2M\ov r}\Big),}
and the horizon at $r=2M$ is a null hypersurface.\pan
Let $\cN$ be a null hypersurface with normal $l$. A vector
$t$ tangent to $\cN$ is characterised by $(t,l)=0$.
But since $l^2=0$, the vector $l$ is itself a tangent
vector, i.e.
\eqnn{
l^\mu={dx^\mu\ov d\lambda},\mtxt{where} x^\mu(\lambda)\mtxt{is a null curve on}
\cN.}
Now one can show, that $\nabla_l l^\mu\vert_{\cN}\sim l^\mu$,
which means that $x^\mu(\lambda)$ is a geodesic with tangent
$l$. The function $g$ can be chosen such that $\nabla_l l=0$,
i.e. so that $\lam$ is an affine parameter. 
A null hypersurface $\cN$ is a \textit{Killing horizon} of a Killing
field $K$ if $K$ is normal to $\cN$.\pan
Let $l$ be normal to $\cN$ such that $\nabla_l l=0$.
Then, since on the Killing horizon $K=fl$
for some function $f$, it follows that
\eqnl{
\nabla_K K^\mu=fl^\nu\nabla_\nu(fl^\mu)=fl^\mu l^\nu\pa_\nu f
=(\nabla_K\log \vert f\vert)K^\mu\equiv \kappa K^\mu\mtxt{on}
\cN.}{surfgrav}
One can show, that the \textit{surface gravity} 
$\kappa=\ha\nabla_K\log f^2$ is constant on orbits of $K$.
If $\kappa\neq 0$, then $\cN$ is a bifurcate
Killing horizon of $K$ with bifurcation $2$-sphere $B$.
In this non-degenerate case $\kappa^2$ is constant on $\cN$.
For example, for the Killing field in Rindler
spacetime \refs{kill} $\nabla_K K=\pm\kappa K$
on the Killing horizon and the bifurcation 'sphere' is at
$\rho=0$.
If $\cN$ is a Killing horizon of $K$ with
surface gravity $\kappa$, then it is also a Killing
horizon of $cK$ with surface gravity $c^2\kappa$.
Thus the surface gravity depends on the normalisation
of $K$. For asymptotically flat spacetimes there
is the natural normalisation
$K^2\to 1$ and $K$ future directed as $r\to \infty$.
With this normalisation the surface gravity is the acceleration of 
a static particle near the horizon as measured
at spatial infinity.\pan
A Killing field is uniquely determined by its value and
the value of its derivative $F_{\mu\nu}=\nabla_{[\mu}K_{\nu]}$
at any point $p\in M$. At the bifurcation point $p$ of a bifurcate
Killing horizon $K$ vanishes and hence is
determined by $F_{\mu\nu}(p)$. In two dimensions $F_{\mu\nu}(p)$
is unique up to scaling. The infinitesimal action of the isometries 
$\al_t$ generated by $K$ takes a vector $v^\mu$ at $p$ into
\eqnl{
L_Kv^\mu=F^\mu_{\;\nu}v^\nu.}{infboost}
The nature of this map on $T_p$ depends upon the signature
of the metric. For Riemannian signature it is an infinitesimal
rotation and the orbits of $\al_t$ are closed
with a certain period. For Lorentz signature \refs{infboost}
is an infinitesimal Lorentz boost and the orbits of
$\al_t$ have the same structure as in the Rindler case.
A similar analysis applies to higher dimensions.\pan
\pan
The Rindler wedge $R$ is globally hyperbolic with
Cauchy hypersurface $\Sigma_R$ (see fig. \refs{rindler1}). Thus it
may be viewed as a spacetime in its own right,
and we may construct a quantum field theory
on it. When we do that, we obtain a remarkable
conclusion, namely that the standard Minkowski vacuum 
$\Omega_M$ corresponds to a thermal state
in the new construction. This means, that an
accelerated observer will feel himself to be
immersed in a thermal bath of particles with
temperature proportional to his acceleration $a$
\cite{Unruh},
\eqnn{ 
kT=\hbar a/2\pi c.}
The noise along a
hyperbola is greater than that along a geodesic, and
this excess noise excites the Rindler detector:
A uniformly accelerated detector in its ground state
may jump spontaneously to an excited state.
Note that the temperature tends to zero when
$\hbar$ tends to zero. Such a radiation has non-zero
entropy. Since the use of an accelerated
frame seems to be unrelated to any statistical
average, the appearance of a non-vanishing entropy 
is rather puzzling.
The Unruh effect shows, that at the quantum level
there is a deep relation between the theory of
relativity and the theory of fluctuations associated
with states of thermal equilibrium, two major aspects
of Einstein's work: The distinction between quantum
zero-point and thermal fluctuations is not an invariant one,
but depends on the motion of the observer.
Note that the temperature is proportional to the acceleration
$a$ of the observer. Since $a=1/\rho$ this means that
$T\rho=\hbox{const}\Longleftrightarrow T\sqrt{g_{00}}=\hbox{const.}$
This is just the \textit{Tolman-Ehrenfest relation} \cite{tolman}
for the temperature in a fluid in hydrostatic equilibrium
in a gravitational field. The factor $\sqrt{g_{00}}$ guarantees
that no work can be gained by transferring radiation
between two regions at different gravitational potentials. 

Let us calculate the number of 'Rindler-particles'
in Minkowski vacuum.
To simplify the analysis, we consider a 
zero-mass scalar field in
two-dimensional Minkowski space.
In the Heisenberg picture, the expansions in terms
of annihilation and creation operators are
\eqnn{
\phi=\int dk\Big(a_k u_k+h.c.\Big),
\mtxt{where}u_k={1\ov\sqrt{4\pi\om}}e^{-i\om x^0+ikx^3},\quad
\omega=\vert k\vert}
and
\eqnn{
\phi=\int dp
\Big(b_p v_p+h.c.\Big),\mtxt{where}
v_p={1\ov \sqrt{4\pi\eps}}\rho^{ip/\kappa}\,e^{-i\eps t},\quad
\;\;\eps=\vert p\vert.}
The $\beta$-coefficients are found to be
\eqnn{
\beta(p,k)=-(\bar u_k,v_p)=
{1\ov 4\pi}\intl_0^\infty \Big(\sqrt{\om\ov\eps}-\sqrt{\eps\ov\om}
{1\ov\kappa\rho}\Big)
e^{ik\rho}\rho^{ip}d\rho,}
where we have evaluated the time-independent 'scalar-product' 
at $t=0$ for which $x^0=0$.
Using the formula
\eqnl{
\intl_0^\infty dx\, x^{\nu-1}e^{-(\al+i\beta)x}=
\Gamma(\nu)(\al^2+\beta^2)^{-\nu/2}e^{-i\nu\arctan(\beta/\al)}}{integral}
we arrive at
\eqnn{
\beta(p,k)=-{\Gamma(ip/\kappa)\ov 4\pi\kappa}\,
\om^{-i p/\kappa}\Big(\sqrt{\eps\ov \om}\pm{p\ov\sqrt{\eps\om}}
\Big)e^{\mp \pi p/2\kappa}\mtxt{for}{k\ov\om}=\pm 1,}
or at
\eqnn{
\vert \beta(p,k)\vert^2={1\ov 2\pi\kappa\om}{1\ov e^{2\pi\eps/\kappa}-1}.}
The Minkowski spacetime vacuum is characterised by
$a_k\Omega_M=0\mtxt{for all}k$.
Assuming that this is the state of the system, the
expectation value of the occupation number as 
defined by the Rindler observer,
$n_p\equiv b^\dagger_pb_p$, is found to be
\eqnl{
\big(\Omega_M,n_p\Omega_M\big)=
\int dk\vert\beta(p,k)\vert^2=
\hbox{volume}\times
{1\ov e^{2\pi \eps/\kappa}-1},}{rindlerequ}
Thus for an accelerated observer the quantum field
seems to be in an equilibrium state with temperature
proportional to $T=\kappa/2\pi=a/2\pi$.
An observer with $a=10^{21}$cm/sec$^2$ feels a
temperature $T\sim 1^0K$.
Since $T$ tends to zero as $\rho\to\infty$
the Hawking temperature (i.e. temperature as measured
at spatial $\infty$) is actually zero. This is expected,
since there is nothing inside which could radiate.
But for a black hole
$T_{local}\to T_H$ at infinity
and the black hole must radiate at this temperature.\pan
Let us finally see, how the (massless)
Feynman-Green function in Minkowski spacetime,
\eqnn{
i\Delta_F(x,x^\pr)=\la 0\vert T\big(\phi(x)\phi(x^\pr)\big)\vert 0\ra
={i\ov 4\pi^2}{1\ov (x-x^\pr)^2-i\eps},}
appears to an accelerated observer.
Let $x=(t,\rho)$ and $x^\pr=(t^\pr,\rho)$ be two events on the world line
of an accelerated observer. Since the invariant
distance of these two events is $2\rho\sinh{\kappa\ov 2}(t-t^\pr)$, 
one arrives at the following spectral representation
of the Feynman-propagator as seen by this
observer
\eqnl{\Delta_F(x,x^\pr)={1\ov (2\pi)^4}({\kappa\ov \rho})^2
\int d^4p\,  e^{-iE(t-t^\pr)}\Big(
{1\ov p^2+i\eps}-2\pi i{\delta(p^2)\ov e^{\beta\vert E\vert}-1}\Big).}{therm}
This is the finite temperature propagator. It follows,
that atoms dragged along the world
line find their excited levels populated as predicted
by a temperature $\beta^{-1}=a/2\pi$.
\section{The Stress-Energy Tensor}
Semiclassically one would expect that
back-reaction is described by the 'semiclassical
Einstein equation'
\eqnn{
G_{\mu\nu}=8\pi G \la T_{\mu\nu}\ra,}
where the right-hand side contains the expectation
value of the energy-momentum tensor of the relevant
quantised field in the chosen state. If the
characteristic curvature radius $L$ in a region
of spacetime is much greater then the Planck
length $l_{pl}$, then in the calculation
of $\la T_{\mu\nu}\ra$ one can expand
in the small parameter $\eps=(l_{pl}/L)^2$
and retain only the terms up to first order in 
$\eps$ (one-loop approximation). The term of order
$\eps$, containing a factor $\hbar$, represents
the main quantum correction to the classical result.
In the one-loop approximation or free fields the contributions
of all fields to $\la T_{\mu\nu}\ra$ are additive and thus
can be studied independently.

The difficulties with defining $\la T_{\mu\nu}\ra=\om(T_{\mu\nu})$
are present already in Minkowski spacetime. The divergences
are due to the vacuum zero-fluctuations. 
The methods of extracting a finite, physically
meaningful part, known as renormalisation procedures,
were extensively discussed in the literature
\cite{enmom}. 
A simple cure for this difficulty
is (for free fields) the \textit{normal ordering} prescription.
We first consider the ill-defined object $\phi^2(x)$, 
which is part of the stress-energy tensor. We may
split the points and consider first the object
$\om(\phi(x)\phi(y))$ which solves the Klein-Gordon equation. 
This bi-distribution makes perfectly good sense. 
For physically reasonable states $\om$ in the Fock space
(e.g. states with a finite number of particles)
the singular behaviour of this bi-distribution is the
same as that belonging to the vacuum state, $\om_0\big(\phi(x)\phi(y)\big).$
For such states the difference
\eqnn{
F(x,y)=\om\big(\phi(x)\phi(y)\big)-\om_0\big(\phi(x)\phi(y)\big)}
is a smooth function of its arguments. Hence, after performing
this 'vacuum subtraction' the coincidence limit may be taken.
We then define
\eqnn{
\om\big(\phi^2(x)\big)=\lim_{x\to y}F(x,y).}
The same prescription can be used for the stress-energy
tensor. We define
\eqnl{
\om\big(T_{\mu\nu}(x)\big)=
\lim_{x\to x^\pr}D_{\mu\nu^\pr}F(x,x^\pr),\quad
D_{\mu\nu^\pr}=
\pa_\mu\pa_{\nu^\pr}-\ha g_{\mu\nu}
\big[\pa_\al\pa^{\al^\pr}-m^2\big]\Big).}{enmomm}
In curved spacetime some restrictions should be expected on
the class of states on which $\la T_{\mu\nu}\ra$
can be defined this way. The \textit{Hadamard
condition} provides a restriction of exactly this
sort of states.\pan
Although \refs{enmomm} is not a physical 
definition of expectation values
of the stress-energy tensor itself
(no preferred vacuum state, vacuum polarisation),
it sensibly defines 
the \textit{differences} of the expected stress energy 
between two states.
In the absence of an obvious prescription it is useful to take an
axiomatic approach. Wald showed that a renormalised
stress tensor satisfying certain reasonable physical
requirements is essentially unique
\cite{waldaxiom}. Its ambiguity
can be absorbed into redefinitions of the coupling
constants in the (generalised) gravitational field equation. Wald's
requirements are:\pan
\textbf{Consistency:}
Whenever $\om_1(\phi(x)\phi(y))-\om_2(\phi(x)\phi(y))$
is a smooth function, then $\om_1(T_{\mu\nu})-\om_2(T_{\mu\nu})$
is well-defined and should be given by the above 
'point-splitting' prescription.\pan
\textbf{Conservation:}
There is a regularisation which respects the diffeomorphism
invariance, so that $\nabla_\nu T^{\mu\nu}=0$
holds. This property
is needed for consistency of Einstein's gravitational
field equation.\pan
\textbf{Normalisation:} In Minkowski spacetime, 
we have $(\Omega_M,T_{\mu\nu}\Omega_M)=0.$
\pan
\textbf{Causality:}
For a fixed in-state in an asymptotically static
spacetime $\om_{in}\big(T_{\mu\nu}(x)\big)$ is
independent of variations of $g_{\mu\nu}$ outside
the past light cone of $x$. For a fixed out-state,
$\om_{out}\big(T_{\mu\nu}\big)$ is independent of 
metric variations outside the future light cone of $x$.
\pan
The Causality axiom can be replaced by a locality property, 
which does not assume an asymptotically static spacetime.
The first and last properties are the key ones,
since they uniquely determine the expected stress-energy
tensor up to the addition of local curvature terms:\pan
\textbf{Uniqueness theorem:} Let $T_{\mu\nu}$
and $\tilde T_{\mu\nu}$ be operators on globally
hyperbolic spacetime satisfying
the axioms of Wald. Then the difference
$U_{\mu\nu}=T_{\mu\nu}-\tilde T_{\mu\nu}$
is a multiple of the identity operator,
is conserved, $\nabla_\nu U^{\mu\nu}=0$ and
is a local tensor of the metric. That is, it
depends only on the metric and its derivatives,
via the curvature tensor, at the same point $x$.
As a consequence
$\om(T_{\mu\nu})-\om(\tilde T_{\mu\nu})$
is independent of the state $\om$ and depends only
locally on curvature invariants.
The proofs of these properties are rather simple and
can be found in the standard textbooks.\pan
\textbf{Calculating the stress-energy tensor.}
A 'point-splitting' prescription where one subtracts
from $\om(\phi(x)\phi(y))$ the expectation value
$\om_0(\phi(x)\phi(y))$ in some fixed state $\om_0$
fulfils the consistency requirement, but cannot
fulfil the first and third axiom at the same time.
However, if one subtracts a locally constructed
bi-distribution $H(x,y)$ which satisfies the wave equation,
has a suitable singularity structure and is equal
to $(\Omega_M,\phi(x)\phi(y)\Omega_M)$ in Minkowski spacetime,
then all four properties will be satisfied.\pan
To find a suitable bi-distribution one recalls the
singularity structure \refs{hadamard}
of $\om_2(x,y)$. In Minkowski spacetime and for massless fields
$w=0$ and this suggests that we take the bi-distribution
\eqnn{
H(x,y)={u(x,y)\ov \sigma}+v(x,y)\log\sigma}
For massless fields the resulting stress-energy obeys all
properties listed above (for massive fields a slight modification
is needed). \pan
\textbf{Effective action.}
The classical metric energy momentum
tensor 
\eqnn{
^{cl}T_{\mu\nu}(x)=
{2\ov \sqrt{\vert g\vert}}{\delta S\ov \delta g^{\mu\nu}(x)}}
is symmetric and conserved (for solutions
of the field equation) for a diffeomorphism-invariant
classical action $S$. If we could construct a 
diffeomorphism-invariant \textit{effective action } 
$\Gamma$, whose variation 
with respect to the metric yields an expectation value of
the energy momentum tensor,
\eqnn{
\la T_{\mu\nu}(x)\ra ={2\ov \sqrt{\vert g\vert}}{\delta \Gamma\ov
\delta g^{\mu\nu}(x)},}
then $\la T_{\mu\nu}\ra$ would be conserved by construction. 
There exists a number of procedures for regularising
$\la T_{\mu\nu}\ra$, i.e. dimensional, point-splitting or zeta-function
regularisation, to mention the most popular ones.
Unfortunately the 'divergent' part' of 
$T_{\mu\nu}$ cannot be completely absorbed into the
parameters already present in the theory, i.e.
gravitational and cosmological constant and parameters
of the field theory under investigation. One finds
that one must introduce new, dimensionless parameters.\pan
The regularisation and renormalisation of the
effective action is more transparent. The divergent geometric parts 
of the effective action, $\Gamma=\int \eta \gam_{div}+\Gamma_{finite}$
have in the one-loop approximation the form
\eqnn{
\gam_{div}=A+BR+C(\hbox{Weyl})^2+D\big[(\hbox{Ricci})^2-R^2\big]
+E\nabla^2R+FR^2.}
Only the part containing $A$ and $B$ can be absorbed into
the classical action of gravity. The remaining terms
with dimensionless parameters $C-F$ lead,
upon variation with respect to the metric,
to a $2$-parameter ambiguity in the expression
for $T_{\mu\nu}$.\pan
\textbf{Effective actions and $\la T_{\mu\nu}\ra$ in two dimensions.}
In two dimensions there are less divergent terms
in the effective action. They have the form
$\gamma_{div}=A+BR$.
The last topological term does not
contribute to $T_{\mu\nu}$ and the first one leads to
an ambiguous term $\sim Ag_{\mu\nu}$ in the energy 
momentum tensor.

The symmetric stress-energy tensor has
$3$ components, two of which are (almost) determined
by $T^{\mu\nu}_{\;\; ;\nu}=0$. As independent 
component we choose the trace $T=T^\mu_{\;\mu}$
which is a scalar of dimension $L^{-2}$.
The ambiguities in the reconstruction of $T^{\mu\nu}$
from its trace is most transparent if we choose
isothermal coordinates for which
\eqnn{
ds^2=e^{2\sigma}\Big((dx^0)^2-(dx^1)^2)\Big).}
This is possible in two dimensions.
Introducing null-coordinates
\eqnn{
u=\ha(x^0-x^1)\mtxt{and} v=\ha(x^0+ x^1)\Rightarrow
ds^2=4e^{2\sigma}dudv,}
the non-vanishing Christoffel symbols are
$\Gamma^u_{uu}=2\pa_u\sigma,\;
\Gamma^v_{vv}=2\pa_v\sigma$
and the Ricci scalar reads $R=-2e^{-2\sigma}\pa_u\pa_v\sigma$.
Rewriting the conservation in null-coordinates we obtain
\eqnl{
\pa_u \la T_{vv}\ra+e^{2\sigma}\pa_v\la T\ra=0\mtxt{,}
\pa_v \la T_{uu}\ra+e^{2\sigma}\pa_u\la T\ra=0,}{conslc}
where $T=T^\mu_{\;\mu}=e^{-2\sigma}T_{uv}$.
The trace $\la T\ra$ determines $\la T_{vv}\ra$
up to a function $t_v(v)$ and $\la T_{uu}\ra$ up
to a function $t_u(u)$. These free functions contain
information about the state of the quantum system.\pan
In the case of a classical conformally invariant
field, $^{cl}T^\mu_{\;\mu}=0$. An important feature of
$\la T_{\mu\nu}\ra$ is that its trace does not vanish
any more. This trace-anomaly
is a state-independent local scalar of dimension $L^{-2}$
and hence must be proportional to the Ricci scalar,
\eqnn{
\la T\ra={c\ov 24\pi}R=-{c\ov 12\pi}e^{-2\sigma}\pa_u\pa_v\sigma,}
where $c$ is the \textit{central charge}.
Inserting this trace anomaly into \refs{conslc} 
yields
\eqnl{
\la T_{uu,vv}\ra=-{c\ov 12\pi}e^\sigma \pa^2_{u,v} 
e^{-\sigma}+t_{u,v}\mtxt{and} \la T_{uv}\ra=-
{c\ov 12\pi}\Box_0\sigma.}{enmink}
Formally, the expectation value of the stress-energy
tensor is given by the path integral
\eqnn{
\la T_{\mu\nu}(x)\ra=-{1\ov Z[g]}\int \cd\phi\; 
{2\ov \sqrt{g}}{\delta\ov \delta g^{\mu\nu}}e^{-S[\phi]}
={2\ov \sqrt{g}}{\delta\ov \delta g^{\mu\nu}}\Gamma[\phi],}
where the effective action is given by
\eqnn{
\Gamma[g]=-\log Z[g]=-\log \int \cd\phi\; e^{-S[\phi]}=
\ha\log\det(-\triangle_c)}
and we made the transition to Euclidean spacetime (which is
allowed for the $2d$ models under investigation).
For arbitrary spacetimes the spectrum of $\triangle_c$ is not
known. However, the variation of $\Gamma$ with respect
to $\sigma$ in
$g_{\mu\nu}=e^{2\sigma}\hat g_{\mu\nu}$
is proportional to the expectation value of
the trace of the stress-energy tensor,
\eqnn{
{\delta\Gamma\ov\delta\sigma(x)}=-2g^{\mu\nu}(x){\delta \Gamma
\ov \delta g^{\mu\nu}(x)}=-\sqrt{g}\la T^\mu_{\;\mu}(x)\ra}
and can be calculated for conformally coupled particles
in conformally flat spacetimes. From the conformal
anomaly one can (almost) reconstruct the effective action.
In particular, in two dimensions the
result is the \textit{Polyakov effective action}
\eqnn{
\Gamma[g]-\Gamma[\delta]={c\ov 96\pi}\int \sqrt{g}R{1\ov \triangle}R,}
where the central charge $c$ is $1$ for
uncharged scalars and Dirac fermions
\footnote{see \cite{wipfsachs} for modifications of
this result, for a spacetime with nontrivial topology.}. 
The $\la T_{\mu\nu}\ra$
is found by differentiation with respect to the metric.
The covariant expression is
\eqnl{
\la T_{\mu\nu}\ra={c\ov 48\pi}\Big(2g_{\mu\nu}R-2\nabla_\mu\nabla_\nu S
+\nabla_\mu S\cdot\nabla_\nu S-
\ha g_{\mu\nu}\nabla^\al S\cdot\nabla_\al S\Big),\qquad 
S={1\ov \triangle }R,}{stressen2d}
and in isothermal coordinates this simplifies to \refs{enmink},
as it must be.
This energy-momentum tensor is consistent, conserved
and causality restricts the choice of the 
Green function $1/\triangle$. The ambiguities in
inverting the wave operator in
\refs{stressen2d} shows up in the free
functions $t_{u,v}$. 
A choice of these functions is
equivalent to the choice of a state.\pan
Let us now apply these results
to the $(t,r)$ part of the Schwarzschild
black hole
\eqnn{
ds^2=\al(r)dt^2-{1\ov \al(r)}\,dr^2,\qquad \al(r)=1-{2M\ov r},\qquad
(G=1)}
which we treat as two-dimensional black 
hole\footnote{The resulting energy-momentum tensor is not
identical to the tensor that one gets when one quantises 
only the $s$-modes in the four-dimensional 
Schwarzschild metric \cite{wimu}.}.
We use the 'Regge-Wheeler tortoise coordinate' 
$r_*=r+2M\log\big(r/M-2\big)$,
such that the metric becomes conformally flat,
$ds^2=\al\big(dt^2-dr_*^2\big)$.
and introduce null-coordinates
$2u=t-r_*\mtxt{and}2v=t+r_*.$
Using $\pa_{r_*}=\al \pa_r$ we obtain
for the light-cone components \refs{enmink} 
of the energy momentum tensor
\eqnn{
\la T_{uu,vv}\ra=
-{c\ov 12\pi}\Big({2M\al\ov r^3}+{M^2\ov r^4}\Big)+t_{u,v},\quad
\la T_{uv}\ra=-{c\ov 12\pi}{2M\al\ov r^3}}
or for $\la T_{\mu\nu}\ra$ in the
$x^\mu=(t,r_*)$ coordinate system
\eqnl{
\la T_\mu^{\;\,\nu}\ra=-{cM\ov 24\pi r^4}\pmatrix{
4r+{M\ov\al}&0\cr 0&-{M\ov\al}}
+{1\ov 4\al}\pmatrix{t_u+t_v&t_u-t_v\cr t_v-t_u&-t_u-t_v}.}{2denergy}
The \textit{Boulware state} is the state appropriate
to a vacuum around a static star and contains no radiation at spatial
infinity ${\cal J}^\pm$. Hence $t_u$ and $t_v$
must vanish. This state is singular at the horizon. To see
that, we use regular Kruskal coordinates:
\eqnl{
U=-e^{-u/2M}\mtxt{and}V=e^{v/2M}\mtxt{so that}
ds^2={16M^3\ov r}e^{-r/2M}dUdV.}{aha}
With respect to these coordinates the energy-momentum 
tensor takes the form
\eqnn{
\la T_{UU}\ra=4\big({M\ov U}\big)^2\la T_{uu}\ra,\qquad
\la T_{VV}\ra=4\big({M\ov V}\big)^2\la T_{vv}\ra\mtxt{and}
\la T_{UV}\ra=-4{M^2\ov UV}\la T_{uv}\ra.}
For the Boulware vacuum
$t_u=t_v=0$ and $\la \dots\ra$ is singular at the past horizon at $V=0$
and future horizon  at $U=0$.
The component $\la T_{UU}\ra$ is regular at the future horizon
if $M^2t_u=c/192\pi$ and $\la T_{VV}\ra$ is regular at
the past horizon if $M^2t_v=c/192\pi$. 
The state regular at both horizons is the 
\textit{Israel-Hartle-Hawking state}. In this state
the asymptotic form of the energy-momentum tensor is
\eqnl{
\la 0_{HH}\vert T^\mu_{\;\nu}\vert 0_{HH}\ra\sim
{c\ov 384\pi M^2}\pmatrix{1&0 \cr 0&-1}={c\pi\ov 6}(kT)^2
\pmatrix{1&0 \cr 0& -1}}{hartlehawing}
with $T=1/8\pi kM= \kappa/2\pi k$. This is the stress-tensor
of a \textit{bath} of thermal radiation at temperature $T$.
Finally, demanding that energy-momentum is regular at
the future horizon and that there is no incoming radiation, i.e.
$M^2 t_u=c/192\pi$ and $t_v=0$, results in
\eqnl{
\la 0_{U}\vert T^\mu_{\;\nu}\vert 0_{U}\ra\sim
{c\ov 768\pi M^2}\pmatrix{1&1 \cr -1&-1}={c\pi\ov 12}(kT)^2
\pmatrix{1&1 \cr - 1& -1}}{unruh}
The \textit{Unruh state} is regular on the future
horizon and singular at the past horizon. It describes
the Hawking evaporation process with only outward
flux of thermal radiation.\pan
\textbf{Euclidean Black Holes.}
The most elegant and powerful derivation of the Hawking
radiation involves an adaption of the techniques
due to Kubo to show that the Feynman propagator
for a spacetime with stationary black hole satisfies
the KMS condition.
Consider a system with time-independent
Hamiltonian $H$.
The time evolution of an observable $A$ in the Heisenberg picture is
$A(z)=e^{izH}Ae^{-izH}$,
where $z=t+i\tau$ is complex time. For $\tau=0$ ($t=0$) it is
the time-evolution in a static spacetime with
Lorentzian (Euclidean) signature.
If $\exp(-\beta H),\beta>0$ is trace class, one can define
the equilibrium state of temperature $T=1/\beta$:
\eqnl{
\la A\ra_\beta={1\ov Z}\tr e^{-\beta H}A,\qquad Z=\tr e^{-\beta H}.}
{therev}
Let us introduce the finite temperature correlation functions
\eqngr{
G^\beta_+(z,\vx,\vy)&=&\la \phi(z,\vx)\phi(0,\vy)\ra_\beta
={1\ov Z}\tr\Big(e^{i(z+i\beta) H}\phi(0,\vx)e^{-izH}\phi(0,\vy)\Big)}
{G^\beta_-(z,\vx,\vy)&=&\la \phi(0,\vy)\phi(z,\vx)\ra_\beta
={1\ov Z}\tr\Big(\phi(0,\vy)e^{izH}\phi(0,\vx)e^{-i(z-i\beta)H}\Big)}
We have used the cyclicity under the trace. 
Both exponents in $G_+$ have negative real parts
if $-\beta<\tau<0$; for $G_-$ the condition reads $0<\tau<\beta$.
Therefore, these formulae define holomorphic functions
in those respective strips with boundary values 
$G_\pm^\beta(t,\vx,\vy)$.
It follows immediately, that
\eqnl{
G_-^\beta(z,\vx,\vy)=G_+^\beta(z-i\beta,\vx,\vy)}{KMS1}
which is the KMS-condition \refs{KMS1}.
This condition is now accepted
as a definition of 'thermal equilibrium at temperature
$1/\beta$'. \pan
So far the analytic functions $G_\pm$ have been defined
in disjoint, adjacent strips in the complex time plane.
The KMS-condition states that one of these is the
translate of the other and this allows us to define
a periodic function throughout the complex plane, with
the possible exception of the lines $\tau=\Im(z)=n\beta$.
Because of locality $\phi(x)$ and $\phi(y)$ commute
for space-like separated events and 
\eqnn{
[\phi(t,\vx),\phi(0,\vy)]=0\mtxt{for}t\in I\subset R.}
Then the boundary values of $G_\pm^\beta$
coincide on $I$ and we conclude (by the edge-of-the-wedge 
theorem) that they are restrictions of a single holomorphic,
periodic function, ${\cal G}^\beta(z,\vx,\vy)$, defined in 
a connected region in the complex time plane except parts
of the lines $\tau=n\beta$. \pan
With these preparations we are now ready to show
that the Green function in Schwarzschild spacetime
satisfies the KMS-condition. 
Starting with the analytically continued Schwarzschild metric
\eqnn{
ds^2=\al dz^2-{1\ov \al}dr^2-r^2d\Omega^2,\qquad \al=1-2M/r,\quad
z=t+i\tau,}
we perform the same
coordinate transformation to (complex) Kruskal coordinates
as we did for the Lorentzian solution:
\eqnn{
Z=V+U=2e^{r_*/4M}\sinh{z\ov 4M}\mtxt{and}
X=V-U=2e^{r_*/4M}\cosh{z\ov 4M}.}
The line element reads
\eqnn{
ds^2={16M^3\ov r}e^{-r/2M}\Big(dZ^2-dX^2\Big)-r^2d\Omega^2}
and the Killing field takes the form
\eqnn{
K=\pa_z={1\ov 4M}\Big(Z\pa_X+X\pa_Z\Big)=
{1\ov 4M}\Big(V\pa_V-U\pa_U\Big).}
Setting $Z=T+i{\cal T}$ the orbits of $K$ are
\eqnn{
\pmatrix{T\cr X}=2e^{r_*/4M}\pmatrix{\sinh t/4M\cr \cosh t/4M}
\mtxt{and}
\pmatrix{{\cal T}\cr X}=2e^{r_*/4M}\pmatrix{\sin \tau/4M\cr 
\cos \tau/4M},}
in the Lorentzian and Euclidean slices, respectively. 
As expected from the general properties of
bifurcation spheres, these are Lorentz-boosts and 
rotations, respectively. Since the Euclidean slice
is periodic in $\tau$, the
analytic Green function $\cG(z=t+i\tau,\vx,\vy)$ 
is periodic in imaginary time $\tau$ with period $8\pi M$. 
This corresponds to a temperature $T=1/8\pi M$, the Hawking 
temperature.\pan 
The vector field (with affine parametrisation)
normal to the Killing horizon $\cN$ (the past
and future horizons) is $l=\pa_V$ on the future horizon and
$l=\pa_U$ on the past horizon.
It follows that the surface gravity $\kappa$ (see \refs{surfgrav})
is $1/4M$ on the future horizon and 
$-1/2M$ on the past horizon.\pan
\textbf{Energy-momentum tensor near a black hole.}
In any vacuum spacetime $R_{\mu\nu}$ vanishes and so do
the two local curvature terms which enter the formula for
$T_{\mu\nu}$ with undetermined coefficients. 
Hence $T_{\mu\nu}$ is well-defined in the Schwarzschild spacetime.
The symmetry of $\la T^\mu_{\;\nu}\ra$ due to the
$SO(3)$ symmetry of the spacetime of a non-rotating black hole
and the conservation $\nabla_\nu\la T^{\mu\nu}\ra$
reduce the number of independent components
of $\la T^\mu_{\;\nu}\ra$. Christensen
and Fulling \cite{crfu} showed that in the coordinates
$(t,r_*,\theta,\phi)$ the tensor is block diagonal.
The $(t,r_*)$ part admits the representation
\eqnl{
\la T^\mu_{\;\nu}\ra=\pmatrix{
{T\ov 2}-{H+G\ov \al r^2}-2\Theta&0\cr
0&{H+G\ov \al r^2}}
+{W\ov 4\pi\al r^2}\pmatrix{1&-1\cr 1&-1}
+{N\ov \al r^2}\pmatrix{-1&0\cr 0&1}}{bhten1}
and the $(\theta,\phi)$-part has the form
\eqnl{
\la T^\mu_{\;\nu}\ra=\big({T\ov 4}+\Theta\big)\pmatrix{1&0\cr 0&1}.}{bhten2}
Here $N$ and $W$ are two constants  
and
\eqngr{
\al(r)&=&\Big(1-{2M\ov r}\Big),\;\quad
T(r)=\la T^\mu_{\;\mu}\ra,\quad\;
\Theta(r)=\la T^\theta_{\;\theta}\ra-{1\ov 4}T(r)}
{H(r)&=&\ha \intl_{2M}^r\big(r^\pr\!-\!M\big)T(r^\pr)dr^\pr,\quad
G(r)=2\intl_{2M}^r\big(r^\pr\!-\!3M)\Theta(r^\pr)dr^\pr.}
The energy-momentum tensor is characterised
unambiguously by fixing two functions $T(r),\Theta(r)$
and two constants $N,W$. The constant $W$ gives the intensity
of radiation of the black hole at infinity and $N$
vanishes if the state is regular on the
future horizon. \pan
The radiation intensity $W$ is non-vanishing only in
the \textit{Unruh vacuum}. It has been calculated for the
massless scalar field $(s=0)$, two-components neutrino
field $(s=1/2)$, electromagnetic field $(s=1)$
and gravitational field $(s=2)$ by Page and Elster \cite{pel}:\pan
\centerline{
\begin{tabular}{l|l|l|l}
$M^2W_0$&$M^2W_{1/2}$&$M^2W_1$&$M^2W_2$\\ \hline
  & & & \\
$7.4\cdot 10^{-5}$&$8.2\cdot 10^{-5}$&$3.3\cdot 10^{-5}$&$0.4\cdot
10^{-5}$\\
\end{tabular}}\pan
The coefficient $N$ vanishes for the Unruh and 
Israel-Hartle-Hawking states.\pan
The calculation of the functions in (\ref{bhten1},\ref{bhten2})
meets technical difficulties connected with the fact that
solutions of the radial mode equation (see below) are not
expressed through known transcendental functions and,
consequently, one needs to carry out renormalisation
in divergent integrals within the framework of numerical
methods. The results for $\la T^t_{\;t}\ra$ and $\la T^r_{\;r}\ra$
for the Israel-Hartle-Hawking and the Unruh states 
have been calculated by Howard/Candelas and Elster \cite{hoca}.\pan
In the Hartle-Hawking state the Kruskal coordinate components
of $\la T_{\mu\nu}\ra$ near the horizon are found to be of order
$1/M^4$. The energy flux into the black hole
is negative, as it must be since the 'Hartle-Hawking vacuum'
is time independent and the energy flux at future infinity
is positive. This is possible since $\la T_{\mu\nu}\ra$
need not satisfy the energy conditions.\pan
\textbf{$s$-wave contribution to $\la T_{\mu\nu}\ra$.}
The covariant perturbation theory for the $4d$
effective action $\Gamma$ as developed in \cite{efac}
is very involved for concrete calculations. Here we
shall simplify the problem by considering $s$-modes of
a minimally coupled massless scalar field propagating
in an arbitrary (possibly time-dependent)
spherically symmetric four-dimensional spacetime.
The easiest way to perform this task is to compute the contribution
of these modes to the effective action. We choose
adapted coordinates for which the Euclidean metric takes the form
\eqnn{
ds^2=\gam_{ab}(x^a)\,dx^adx^b+\Omega^2(x^a)\om_{ij}dx^idx^j,}
where the last term is the metric on $S^2$. Now one
can expand the (scalar) matter field into spherical
harmonics. For $s$-waves, $\phi=\phi(x^a)$,
the action for the coupled gravitational and scalar field is
\eqnn{
S=-{1\ov 4}\int\big[\Omega^2\;^\gam\cR+\;^\om\cR+2(\nabla\Omega)^2\big]
\sqrt{\gam}d^2x+2\pi\int \Omega^2(\nabla\phi)^2\sqrt{\gam}d^2x,}
where $^\gam\cR$ is the scalar curvature of the $2d$ space
metric $\gam_{ab}$, $^\om\cR=2$ is the scalar curvature of $S^2$
and $(\nabla\Omega)^2=\gam^{ab}\pa_a\Omega\pa_b\Omega.$
The purely gravitational part of the action is almost
the action belonging to $2d$ dilatonic gravity with two
exceptions: first, the numerical coefficient in front
of $(\nabla\Omega)^2$ is different and second, the action
is not invariant under Weyl transformation due to the
$^\om\cR$ term. The action is quite different from the
actions usually considered in $2d$ (string-inspired)
field theories, because of the unusual coupling of $\phi$
to the dilaton field $\Omega$. Choosing isothermal
coordinates, $\gam_{ab}=e^{2\sigma}\gam^f_{ab}$,
where $\gam^f_{ab}$ is the metric of the flat $2d$ space,
one arrives with $\zeta$-function methods 
at the following exact result for the effective action 
for the $s$-modes \cite{wimu}
\eqngrr{
\Gamma_s&=&^{(n)}\Gamma_s+^{(i)}\Gamma}
{^{(n)}\Gamma[\sigma,\Omega]&=&{1\ov 8\pi}\int\Big(
{1\ov 12}{^\gam\cR}{1\ov\triangle_\gam}{^\gam\cR}-{\triangle_\gam\Omega\ov
\Omega}{1\ov \triangle_\gam}{^\gam\cR}\Big)\sqrt{\gam}d^2x}
{^{(i)}\Gamma[\Omega]&=&\Gamma_s[\sigma=0,\Omega]=
\ha\log\det\Big(-\triangle_f+{\triangle_f\Omega\ov\Omega}\Big).}
The second contribution $^{(i)}\Gamma$ is invariant
under $2d$ Weyl transformation, whereas the first one
is not. Unfortunately, the determinant cannot be calculated
exactly and one must resort to some perturbation expansion.
For details I refer to \cite{wimu}. Ignoring backscattering
one finds
\eqnn{
^{(1)}\Gamma={1\ov 8\pi}\int\Big(
{1\ov 12}{^\gam\cR}{1\ov\triangle_\gam}{^\gam\cR}-{\triangle_\gam\Omega\ov
\Omega}\times\Big[1+\log{\triangle_\gam\Omega\ov
\mu^2\Omega}\Big]\Big)\sqrt{\gam}d^2x.}
Due to backscattering one needs to add the following term:
\eqnn{
^{(2)}\Gamma=-{\xi\ov 12\cdot 8\pi}\int\Big({^\gam\cR}
{1\ov\triangle_\gam}{^\gam\cR}+\hbox{local terms}\Big)\sqrt{\gam}d^2x,}
where $\xi\sim 0.9$. From the action $\Gamma_2=^{(1)}\!\Gamma+
^{(2)}\!\Gamma$ one
obtains $\la T_{\mu\nu}\ra$ by variation with respect
to the metric. To get the flux of the Hawking radiation
we need to continue back to Lorentzian spacetime by changing
the signs in the appropriate places. According to
\cite{efac} we arrive at the in-vacuum energy-momentum
tensor by replacing $-1/\triangle$ by the retarded
Green function. Neglecting backscattering, the luminosity 
of the black hole is found to be
\eqnn{
L=-{\pi\ov 12}{1\ov (8\pi M)^2}.}
This coincides with the total $s$-wave flux of the
Hawking radiation obtained with other methods \cite{Birrell}
without taking backscattering effects into account.
With backscattering, the Hawking radiation is modified
and compares well with that obtained by other means
\cite{Simkin}.
\section{Wave equation in Schwarzschild spacetime}
We study the classical wave propagation of a Klein-Gordon
scalar field in fig.\ref{confdiag}.
\begin{figure}[ht]
\begin{minipage}[t]{15cm}
\centerline{\epsfysize=8 cm\epsffile{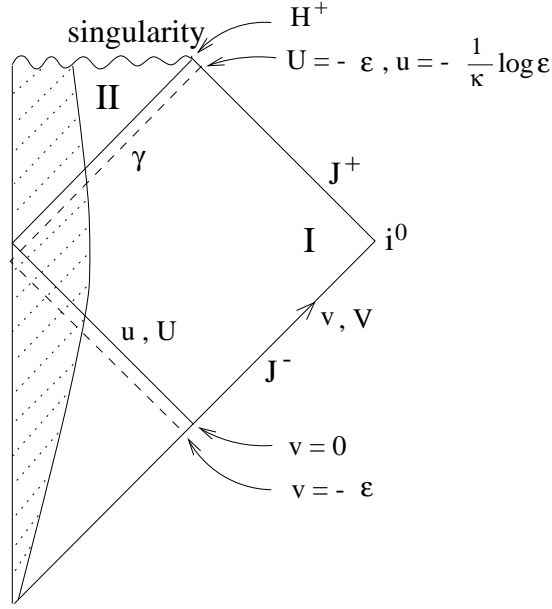}}
\caption{\label{confdiag}\textsl{The propagation
of particles in the geometric optics approximation.}}
\end{minipage}
\end{figure}
At late times, one expects that
every solution will propagate into the black hole
region $II$ and/or propagate to $\cJ^+$.\pan
In the spherically symmetric spacetime we may set
\eqnn{
\phi={f(t,r)\ov r}Y_{lm}e^{-i\om t}}
and the wave equation $(\Box+m^2)\phi=0$
reduces to the radial equation
\eqnl{
{\pa^2f\ov \pa t^2}-{\pa^2f\ov \pa r^2_*}-V(r_*)f=0,
\quad V(r_*)=\Big(1-{2M\ov r}\Big)
\Big({2M\ov r^3}+{l(l+1)\ov r^2}+m^2\Big),}{radialeq}
where $M$ is the mass of the black hole and $m$ that
of the Klein-Gordon field.
As $r_*\to -\infty$ (i.e. $r\to 2M$) the potential falls
off exponentially, $V\sim \exp(r_*/2M)$, and as
$r_*\to\infty$ the potential behaves as $\sim m^2-2Mm^2/r_*$ in
the massive case and $\sim l(l+1)/r^2$ in the massless case.
In the asymptotic region $r\to\infty$ this equation
possesses outgoing solution $\sim e^{i\om r_*}$ and
ingoing solutions $\sim e^{-i\om r_*}$. In terms of
the null-coordinates the asymptotic solutions look like
\eqnl{
f_\om^{out}\sim e^{-2i\om u}\mtxt{and}
f_\om^{in}\sim e^{-2i\om v}.}{inout}
Consider a geometric optics approximation in which
a particle's world line is a null ray, $\gam$, of constant
phase $u$ and trace this ray backwards in time from
$\cJ^+$. The later it reaches $\cJ^+$ the closer it
must approach $H^+$. As $t\to\infty$ the ray $\gam$
becomes a null geodesic generator $\gam_H$ of $H^+$.
We specify $\gam$ by its affine distance from
$\gam_H$ along an ingoing null geodesic through $H^+$
\begin{figure}[ht]
\begin{minipage}[t]{15cm}
\centerline{\epsfysize=5 cm\epsffile{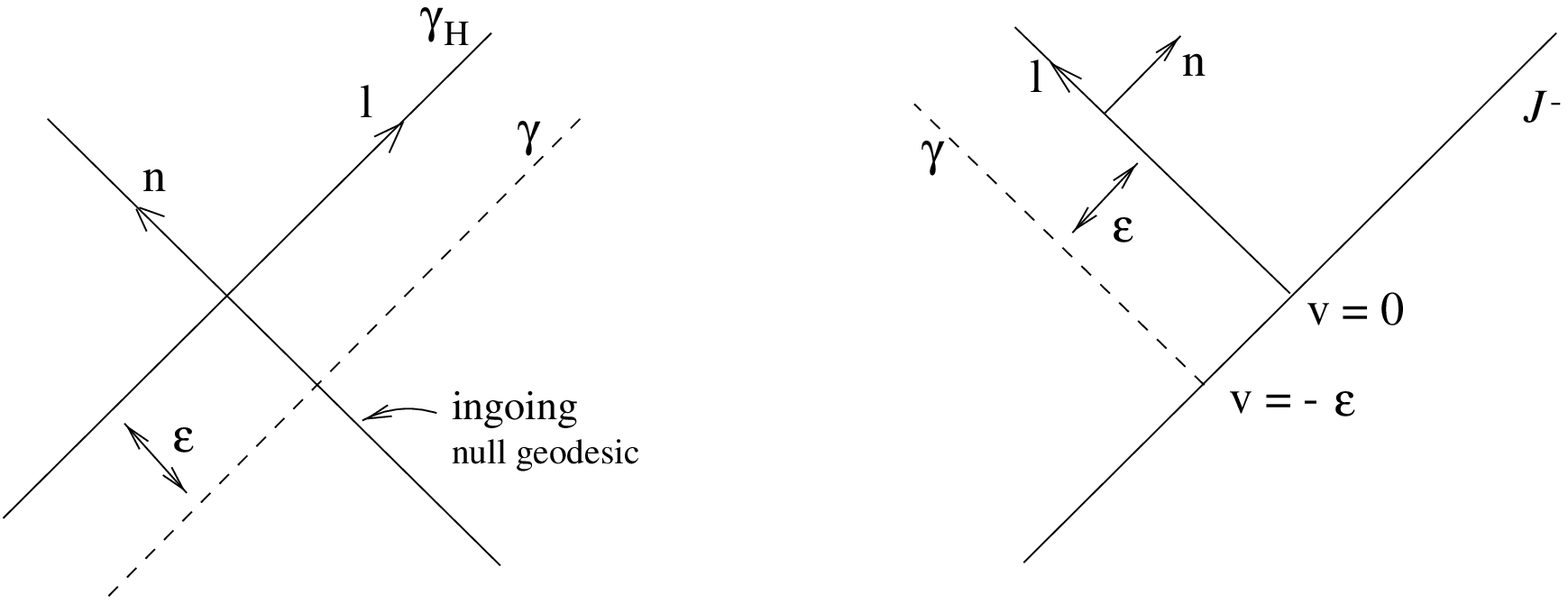}}
\caption{\label{hawkingh}\textsl{ }}
\end{minipage}
\end{figure}
(see fig.\ref{hawkingh}a).
The affine parameter on the ingoing null geodesic is $U$,
so that according to \refs{aha}
\eqnn{
U=-\eps\Rightarrow u=-{1\ov 2\kappa} \log\eps,\quad
f_\om^{out}\sim \exp\Big({i\om\ov \kappa}\log\eps\Big).}
This oscillates rapidly at later times $t$ and
this justifies the geometric optics approximation.
Now we must match $f_\om^{out}$ onto a solution
near $\cJ^-$. In our approximation we just need
to parallel-transport $n$ and $l$ along
the continuation of $\gam_H$ back to $\cJ^-$. 
We choose $v$ such that this continuation meets
$\cJ^-$ at $v=0$. The continuation of $\gam$ will meet
$\cJ^-$ at an affine distance $\eps$ along an outgoing null
geodesic on $\cJ^-$. Since $ds^2=4dudv+\dots$ on $\cJ^-$ 
the coordinate $2v$ is the affine parameter measuring
this distance, so $2v=-\eps$ on $\gam$ and
\eqnn{
f_\om\sim \exp\Big({i\om\ov \kappa}\log(-2v)\Big)\theta(-v),}
where we took into account, that null rays with $v>0$
do not reach $\cJ^+$. Now we take the Fourier transform
\eqnn{
\tilde f_\om(\om^\pr)=\intl_{-\infty}^0e^{2i\om^\pr v} f_\om(v)\,dv
=\ha\intl_0^\infty \tilde v^{i\om/\kappa}e^{-i\om^\pr \tilde v}d\tilde v,\quad
{\om}^\prime >0.}
Using \refs{integral} one sees, that
\eqnn{
\tilde f_\om(\om^\pr)=-e^{\pi\om/\kappa}\tilde f_\om(-\om^\pr)\mtxt{for}
\om^\pr >0.} 
It follows, that a mode of positive frequency $\om$ on $\cJ^+$
matches onto mixed positive and negative frequency modes
on $\cJ^-$. We see, that the Bogolubov coefficients are
related by $\beta_{ij}=-\exp(-\pi\om_i/\kappa)\al_{ij}$.
From the Bogolubov relations \refs{bogrel} one then gets
\eqnl{
\Big(\beta\beta^\dagger\Big)_{ii}={1\ov e^{2\pi\om_i/\kappa}-1}.}{borrel1}
For calculating the late time particle flux through
$\cJ^+$ we need the inverse $\beta$-coefficients, $\beta^\pr=-\beta^t$.
One easily finds, that
$\la N_i\ra_{\cJ^+}=(\beta^{\pr\dagger} \beta^\pr)_{ii}=
(\beta\beta^\dagger)_{ii}$.
This is the Planck-distribution at the Hawking
temperature $T_H=\hbar\kappa/2\pi$.\pan
The detailed form of the potential in $\refs{radialeq}$
is irrelevant in the geometric optics
approximation. But the incoming waves will partially 
scatter off the gravitational field (on the $l$-dependent
potential $V$ in \refs{radialeq})
to become a superposition of incoming and outgoing waves.
The backscattering is a function of $\om$ and the spectrum
is not precisely Planckian. The total luminosity of
the hole is given by 
\eqnl{
L={1\ov 2\pi}\;\sum_{l=0}^\infty \;(2l+1)\intl_0^\infty
d\om \;\om {\Gamma_{\om l}\ov e^{8\pi M\om}-1}.}{luminosity}
A black hole is actually grey, not black. 
The dependency on the angular momentum (and spin) of
the particles resides in the grey-body factor $\Gamma_{\om l}$.
\section{Back-reaction}
The main effect of the quantum field will be a
decrease of $M$ at the rate at which energy is
radiated to infinity by particle creation. 
Since the spacetime is static outside the
collapsing matter, the expected energy current
$J_\mu=\la T_{\mu\nu}\ra K^\nu$ is conserved in that region. 
The calculation showed, that there will be a steady nonzero flux $F$.
In \cite{evap} the contribution of the
different particle species to this flux
has been determined. The contribution of massive particles 
of rest mass $m$ is exponentially small if $m>\kappa$.
Black holes of mass $M>10^{17}$g can only emit neutrinos,
photons and gravitons. Black holes of mass $5\cdot 10^{14}$g$\leq M
\leq 10^{17}$g can also emit electrons and positrons.
Black holes of smaller mass can emit heavier particles.
A non-rotating black hole emits almost as a
body heated to the temperature
\eqnn{
T[^0\hbox{K}]={\hbar\kappa \ov 2\pi c}={\hbar c^3\ov 8\pi G kM}\sim
10^{26}{1\ov M[\hbox{g}]}.}
The deviation from thermal radiation is due to the
frequency dependence of the penetration coefficient
$\Gamma_{s\om l}$. This coefficient is also strongly
spin-dependent,
$\Gamma_{s\om l}\sim \om^{2s+1}$.
As spin increases, the contribution of particles
to the radiation of a non-rotating black hole
diminishes. The distribution of the radiated
particles in different mass-intervals is shown
in the following table:
\vskip 0.5truecm
\pan
\centerline{
\begin{tabular}{l|l|l}
$M\;$[g] & $L\;\big[{\hbox{erg}\ov \hbox{sec}}\big]$ 
& particles radiated \\ \hline
       &                 &  \\
$M>10^{17}$ & $3.5\times 10^{12}\Big({10^{17}g\ov M}\Big)^2$&
$81.4\%\;\,\;\nu_e,\bar\nu_e,\nu_\mu,\bar\nu_\mu$ \\
  & & $16.7\%\;\,\gam\quad 1.9\%\;\,g$ \\ \hline
$10^{17}>M>5\times 10^{14}$ & $ 6.3\times 10^{16}\Big({10^{15}g\ov M}\Big)^2$ & $45\%\;\,\;\nu_e,\bar\nu_e,\nu_\mu,\bar\nu_\mu$ \\
  &  & $9\%\;\,\gam\quad 1\%\;\,g$ \\
  &  & $45\%\;\,e^-,e^+$ \\ \hline
$10^{14}>M>10^{13.5}$ & $ 10^{19}\Big({10^{14}g\ov M}\Big)^2$ &
$48\%\;\,\nu_e,\bar\nu_e,\nu_\mu,\bar\nu_\mu$ \\
 &  & $28\%\;\,e^-,e^+\quad 11\%\;\,\gam$\\
 &  & $1\%\;\,g\quad 12\%\;\,N,\bar N$ \\ \hline
\end{tabular}}
\vskip 0.5truecm
\pan
The following formula describes the rate of mass loss
\eqnl{
-{dM\ov dt}\sim 4\cdot 10^{-5}f\cdot\Big({m_{pl}\ov M}\Big)^2
{m_{pl}\ov t_{pl}}=7.7\cdot 10^{24}f\cdot\Big({1\ov M[\hbox{g}]}\Big)^2
{\hbox{g}\ov \hbox{sec}}={\al\ov M^2}.}{loss}
The contributions of the (massless) particle species
are encoded in $f(M)$. From Page we take
\eqnn{
f=1.02 h(\ha)+0.42 h(1)+0.05 h(2),}
where $h(s)$ is the number or distinct polarisations
of spin-$s$ particles.
The rate equation \refs{loss} is easily integrated to yield
\eqnn{
M(t)=\big(M_0^3-3\al t\big)^{1/3},}
We see that a black hole
radiates all of its mass in a finite time
$\tau\sim M_0^3/3\al$. Inserting for $\al$ yields
\eqnn{
\tau\sim 10^{71}\big({M\ov M_\odot}\big)^3\hbox{sec}.}
If primordial black holes of mass $\sim 5\cdot 10^{14}$g were
produced in the early universe, they would be in
the final stages of evaporation now. Primordial
black hole of smaller mass would have already
evaporated and contributed to the $\gam$-ray background.
See the review of Carr \cite{Carr} for the possibility
of observing quantum explosions of small black holes.\pan
The magnitude of the Kruskal coordinate components of
$\la T_{\mu\nu}\ra_H$ near the black
hole are found to be of order $1/M^4$ in Planck units,
as expected on dimensional grounds. Since the
background curvature is of order $1/M^2$ the quantum
field should only make a small correction to the structure
of the  black hole for $M\gg 1$, or $M\gg 10^{-5}$g. 
\section{Generalisations and Discussion}
In the previous section we have studied the
Hawking effect in the case of the Schwarzschild
black hole. Lets us consider now different
generalisations of this effect and its possible
consequences.\pan
\textbf{Hawking radiation of rotating and charged holes.}
The \textit{Kerr solution} has null-hypersurfaces at
\eqnn{
r=r_\pm=M\pm \sqrt{M^2-a^2},}
where $a=J/M$, which are Killing horizons of the Killing fields
\eqnn{
K_\pm=k+\Omega m=
k+\Big({a\ov r^2_\pm+a^2}\Big)m\qquad k=\pa_t,\quad m=\pa_\phi,}
with surface gravities
\eqnn{
\kappa_\pm={r_\pm-r_\mp\ov 2(r_\pm^2+a^2)}.}
For the extreme Kerr solution with $a^2=M^2$ the 
surface gravity vanishes.

For a Schwarzschild hole the number of particles
per unit time in
the frequency range $\om$ to $\om+d\om$ passing out through
a surface of the sphere is
\eqnn{
{1\ov e^{8\pi M\om}-1}\;{d\om\ov 2\pi}.}
For a Kerr Black hole $\om$ is replaced
by $\om-m\Omega$ in this formula, where $m$ is the
azimuthal quantum number of the spheroidal
harmonics, and $\Omega$ is the angular speed of the
event horizon. Hence, the Planck factor at $J^+$ becomes
\eqnn{
{1\ov e^{2\pi(\om-m\Omega)/\kappa}\pm 1},\qquad
+\,\hbox{fermions},\quad -\hbox{bosons}.}
The emission is stronger for positive $m$ than
for negative $m$. In the boson case the Planck
factor becomes negative when $\om<m\Omega$
and super-radiance occurs: the effect of radiation
amplifies the incoming classical wave with positive $m$.
The result admits the following interpretation: Consider
a rotating black hole enclosed in a mirror-walled
cavity. A scattering of a 'particle' in a super-radiant
mode by the black hole increases the number of
quanta.  After reflection by the mirror, these quanta
are again scattered on the black hole and their number
increases again, and so on. No stationary equilibrium
distribution is possible for such modes. However,
if the size of the cavity is not too large, $r<1/\Omega$,
then the super-radiative modes are absent and equilibrium
is possible.
A related effect is that the rotation of the hole
enhances the emission of particles with higher spins.

For a charged hole with \textit{Reissner-Nordstr\"{o}m metric}
\eqnn{
ds^2=\al(r) dt^2-
{1\ov \al(r)}dr^2+r^2d\Omega^2,\qquad \al(r)=
1-{2M\ov r}+{q^2\ov r^2}}
the event horizon is at $r=r_+=M+\big(M^2-q^2\big)^{1/2}$
and the surface gravity is found to be
\eqnn{
\kappa={1-16\pi^2q^4/A^2\ov 4M},}
where $A=4\pi r_+^2$ is the area of the horizon.
If follows that the presence of the charge depresses
the temperature $kT_H=\kappa/2\pi$ of the hole. 
For an extremal hole with charge $q=M$ or with $a^2=M^2$
the Hawking temperature is zero, whereas the
area is not ($A=4\pi M^2$ for the extreme
Reissner-Nordstr\"{o}m hole). In the laws of black
hole thermodynamics the entropy of
a black hole is $S=A/4$ and hence non-vanishing
for extreme black holes. The formulation
of the third law, namely that $S\to 0$ as $T\to 0$,
is not true for extremal holes\footnote{see the
contribution of Claus Kiefer: the canonical theory
of gravity predicts $S(T\to 0)=0$, whereas
superstring-theory predicts $S(T\to 0)=A/4$.}.
The failure
of the formulation of the third law may not be
too disturbing. There other quantum systems
with a degenerate ground state
for which it fails as well.\pan
\textbf{Loss of Quantum Coherence.}
Consider the behaviour of the quantum field in the
spacetime of a collapse, fig.\ref{collaps}
in which back-reaction effects are not taken
into account. The state of the field at late times
in region $I$, and in particular the flux
of thermal particles reaching infinity, must be described 
by a density matrix. The particles which entered
the black hole at early times are correlated with
the particles in region $I$. 
\begin{figure}[ht]
\begin{minipage}[t]{14cm}
\centerline{\epsfysize=7 cm\epsffile{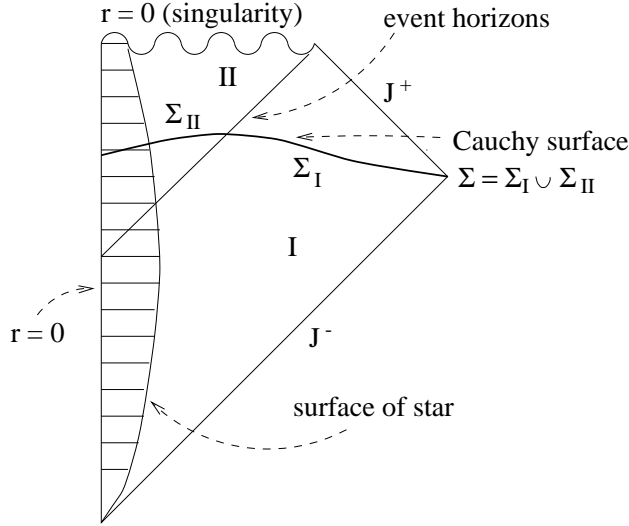}}
\caption{\label{collaps}\textsl{A conformal diagram
of the spacetime resulting from a complete
collapse of a spherical body. The region $II$
lies outside of the chronological past of $J^+$.}}
\end{minipage}
\end{figure}
There is always a
loss of information whenever one performs
an inclusive\footnote{not all commuting observables
are measured} measurement outside the horizon. Such entropy
increase is common to all inclusive measurements in
physics. Perhaps we can understand this situation
better if we recall the resolution of the well-known
question raised by Einstein, Podolsky and Rosen. 
A pure quantum state is defined globally; its coherence
may extend over field variables located at well-separated
points on a space-like surface. \pan
Let us distinguish between the set of out-states
corresponding to particles moving away from the black hole (the visible ones)
and those falling into the hole (the invisible ones).
When one calculates expectation values 
$\la A\ra=\big(\psi,A\psi)$ of operators $A$ 
depending only on the creation and 
annihilation operators belonging to the visible modes,
this expectation value can be written as
$\la A\ra =\tr\rho A$.
In a Fock space construction one can derive an explicit
formula for the density matrix $\rho$ in
terms of the pure state $\psi$. Here it suffices to
sketch the emergence of a mixed state from a pure
one. let $\psi=\psi^I_i\otimes\psi^{II}_j$ be
orthonormal pure states in the big Hilbert space
$\ch=\ch_I\otimes \ch_{II}$. Let us further assume 
that the observable $A$ is the identity in $\ch_{II}$.
Then the expectation value
\eqnn{
(\psi,A\psi)\mtxt{in the pure state}\psi=\sum \al_i\psi_i^I\otimes
\psi_i^{II},\quad
\sum \vert\al_i\vert^2=1}
becomes
\eqnn{
(\psi,A\psi)=\sum_{ij}\bar\al_i\al_j\big(\psi_i^I\otimes \psi^{II}_i,
A\psi^{I}_j\otimes \psi^{II}_j\big)
=\sum p_i(\psi_i^I,A\psi^I_i)=\tr (\rho A),}
where $p_i=\vert\al_i\vert^2$ and $\rho=\sum p_iP_i$.
The $P_i$ are the projectors on the states $\psi^I_i$.
We have used, that the $\psi^{II}_i$ are orthonormal.
Thus, if we are only measuring observables in the region
$I$ outside of the black hole and ignore the information
about the inside, then pure states become indeed mixed
states. For a black hole $\al_i\sim \exp(-\pi\om_i/\kappa)$
(see \refs{borrel1}) and $\rho$ is the thermal state.
As is also clear, for operators $A$ which are not the identity
in $\cH_I$ the expectation values $(\psi,A\psi)$ cannot 
be written as $\tr\rho A$.
\pan
Consider now the spacetime fig.\ref{evap} in which back-reaction
causes the black hole to 'evaporate'. 
\begin{figure}[ht]
\begin{minipage}[t]{14cm}
\centerline{\epsfysize=7 cm\epsffile{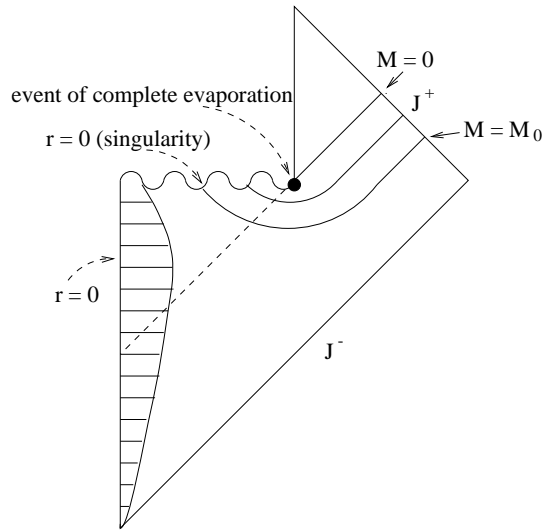}}
\caption{\label{evap}\textsl{A conformal diagram
of a spacetime in which black hole formation and
evaporation occurs. The contour labelled $M=0$
lies at the (retarded) time corresponding
to the final instant of evaporation.}}
\end{minipage}
\end{figure}
The visible
particles propagating to infinity can be described by
a (thermal) density matrix. The particle creation
and scattering will be described by a unitary $S$-matrix,
provided that the invisible particles are represented
in the 'out'-Hilbert space. What happens now when the black
hole disappears from the spacetime? Apparently at late
times, if one takes the 'out'-Hilbert space
to be the Fock space associated with visible particles,
the entire state of the field is mixed. Then  one cannot
describe particle creation and scattering by a unitary
$S$-matrix, since an initial pure state evolved into a
density matrix. This is the phenomenom of \textit{loss of
quantum coherence}. 
What are the possible ways out of this problem?
A complete calculation including all back-reaction
effects might resolve the issue, but even this is
controversial, since the resolution very probably
requires an understanding of the Planck scale physics.
For example, $QFT$ predicts that $T_{loc}\to\infty$
on the horizon of a black hole. This should not
be believed when $T$ reaches the Planck energy.
The quantum aspects of gravity cannot be any longer
ignored and this temperature is then of the order
of the maximum (Hagedorn) temperature of string theory
\footnote{See the contribution of G. 't Hooft.}.\pan
A natural approach to dealing with this situation
is to consider 'toy models', for example in two spacetime
dimensions, in which the semiclassical
analysis could be done. In lower dimensions one adds
a 'dilaton' field to render gravity non-trivial
(this field naturally arises in low energy
string theory). The resulting two-dimensional theories
are dynamically nontrivial and mimic many
features of four-dimensional general relativity:
they possess black-hole solutions, Hawking radiation
and there exist laws of black hole thermodynamics
which are completely analogous to the laws in
four dimensions. Callen et.al \cite{callen} studied the model
\eqnl{
S={1\ov 2\pi}\int d^2x\sqrt{-g}\Big(e^{-2\sigma}\big[
R+4(\nabla\sigma)^2+4\lam^2\big]+\ha(\nabla f)^2\Big),}{callan}
containing a metric field $g_{\mu\nu}$, a dilaton field
$\sigma$ and a matter field $f$. The Hawking radiation
of the $f$-'particles' can be calculated the way
we explained in our two-dimensional model calculations
above. So far these model calculations
have not resolved the problems with the final stage
of the black hole evaporations (the problems
are the same as those with the Liouville
theory at strong-coupling). A further simplification of \refs{callan}
has been discovered by Russo, Susskind and Thorlacius \cite{RST}.
Rather recent calculations seem to indicate\footnote{
see the contribution of C. Kiefer} that information 
is not destroyed, but slowly
released as the black hole decays back to vacuum \cite{strom}.

\end{document}